\documentclass[13pt,letterpaper]{article}

\usepackage[english]{babel}
\usepackage[utf8x]{inputenc}
\usepackage{booktabs}
\usepackage{tabu}
\usepackage[T1]{fontenc}
\usepackage{comment}
\usepackage[letterpaper,top=2.54cm,bottom=2.54cm,left=2.54cm,right=2.54cm,marginparwidth=2.54cm]{geometry}
\usepackage{amsmath}
\usepackage{amssymb}
\usepackage{tipa}
\usepackage{enumitem}
\usepackage{xcolor, soul}
\usepackage{graphicx}
\usepackage{placeins}
\usepackage{outlines}
\usepackage{amssymb}
\usepackage{bbm}
\usepackage{xurl}
\usepackage[numbers,sort&compress]{natbib}

\usepackage{listings,xcolor}

\definecolor{dblue}{rgb}{0.0, 0.0, 0.5}

\definecolor{dgreen}{rgb}{0.0, 0.2, 0.0}

 	\definecolor{cblue}{rgb}{0.0, 0.0, 0.4}

\newcommand{\cblue}[1]{\textcolor{cblue}{#1}}
\definecolor{colorkind}{rgb}{0.47, 0.27, 0.23}
\newcommand{\colorkind}[1]{\textcolor{colorkind}{#1}}
\definecolor{fielddrab}{rgb}{0.0, 0.15, 0.05}
\newcommand{\colorvar}[1]{\textcolor{fielddrab}{#1}}
\usepackage{algorithm}
\usepackage[noend]{algpseudocode}
\usepackage[colorinlistoftodos]{todonotes}
\usepackage[colorlinks=true, allcolors=blue]{hyperref}
\newcommand{\kn}[1]{\textcolor{blue}{\textbf{KN:} #1}}

\newcommand{\gc}[1]{\textcolor{brown}{\textbf{GC:} #1}}
\title{\vspace{-45pt}\textbf{FEDSTR (f\textepsilon dst\textschwa r): Money-In  AI-Out
\\A Decentralized Marketplace for Federated Learning and LLM Training on the NOSTR Protocol  \\
       \large [Proof-of-Concept --- \small Code: \url{https://github.com/ConstantinosNikolakakis/Fedstr}]
}\vspace{-5pt}} 

\renewenvironment{thebibliography}[1]{
  \begin{oldthebibliography}{#1}
    \setlength{\itemsep}{0.115em}
    \setlength{\parskip}{0em}
}
{
  \end{oldthebibliography}
}
\author{Konstantinos E. Nikolakakis\\
\texttt{\small KostisNikolakakis@pm.me} 
\and George Chantzialexiou\\
\texttt{\small george.chantzialexiou@gmail.com} 
\and Dionysis Kalogerias\\
\texttt{\small dionysis.kalogerias@yale.edu}}

\date{}

    \usepackage{ucs}
    \usepackage{amsmath}
    \usepackage{amsfonts}
    \usepackage{amssymb}
    \usepackage{float}
    
    \newfloat{lstfloat}{htbp}{lop}
    \floatname{lstfloat}{Listing}
    \usepackage{xcolor}
    \usepackage{listings}
    \usepackage{pgf}
    \usepackage{pgffor}

    \usepackage[
        style=1,
        skipbelow=\topskip,
        skipabove=\topskip
    ]{mdframed}

    \definecolor{bggray}{rgb}{0.85, 0.85, 0.85}
    \mdfsetup{
        leftmargin = 20pt,
        rightmargin = 20pt,
        backgroundcolor = bggray,
        middlelinecolor = black,
        roundcorner = 15
    }
    \BeforeBeginEnvironment{lstlisting}{\begin{mdframed}\vskip-.5\baselineskip}
    \AfterEndEnvironment{lstlisting}{\end{mdframed}}

\begin{document}
\lstset{
    string=[s]{"}{"},
    stringstyle=\color{dgreen},
    comment=[l]{:},
    commentstyle=\color{dblue},
}

\maketitle
\vspace{-15pt}

\begin{abstract}

The NOSTR is a communication protocol for the social web, based on the w3c websockets standard. Although it is still in its infancy, it is well known as a social media protocol, with thousands of trusted users and multiple user interfaces, offering a unique experience and enormous capabilities. To name a few, the NOSTR applications include but are not limited to direct messaging, file sharing, audio/video streaming, collaborative writing, blogging and data processing through distributed AI directories. In this work, we propose an approach that builds upon the existing protocol structure with end goal a decentralized marketplace for federated learning and LLM training. In this proposed design there are two parties: on one side there are customers who provide a dataset that they want to use for training an AI model. On the other side, there are service providers, who receive (parts of) the dataset, train the AI model, and for a payment as an exchange, they return the optimized AI model. To demonstrate viability, we present a proof-of-concept implementation over public NOSTR relays. The decentralized and censorship resistant features of the NOSTR enable the possibility of designing a fair and open marketplace for training AI models and LLMs.
\end{abstract}


\section{Introduction}


The NOSTR (n\textopeno st\textschwa r) - Notes and Other Stuff Transmitted by Relays~\cite{nostr-protocol} is an open protocol that enables censorship-resistant communication. In fact, it is resilient and does not rely on any trusted central servers. The communication over the NOSTR is tamperproof since it relies on cryptographic keys and signatures. The first successful application of the NOSTR protocol is the decentralized social media with at least $150000$ known trusted users~\cite{nostrband}. The NOSTR has several common structural properties with the competitive open protocol Bluesky~\cite{bluesky_arch,kleppmann2024bluesky}. Bluesky and the NOSTR both aim to solve the problem of decentralized social media, under partial consistency trade-offs~\cite[CAP Theorem]{gilbert2012perspectives,wei2024exploring}. Herein, we focus on the NOSTR protocol because it is the most flexible and safest of the protocols in terms of developer control and future development. 

In parallel with the NOSTR protocol development, distributed AI training has attracted major attention over the last few years, with the Language Model breakthroughs leading the progress. By combining recent advanced tools and breakthroughs from open communication protocols~\cite{nostr-protocol}, bitcoin payments~\cite{poon2016bitcoin} and distributed optimization algorithms~\cite{douillard2023diloco}, we propose a system design that enables open marketplaces for training AI models. Specifically, the NOSTR has already built-in protocol flow for payments through the lightning network~\cite{poon2016bitcoin}, as well as communication flows for a distributed data processing system, which makes it the state of the art protocol for designing a decentralized market for distributed computation and AI directories. Such a network system can mitigate privacy issues, security risks and potentially minimize operational costs of machine learning and language models training by leveraging the structure and capabilities of the NOSTR protocol. We begin by discussing some fundamental aspects of the NOSTR, and the core protocol application designed specifically for the social web.

A prominent challenge for social web applications is designing decentralized social networks through peer-to-peer architectures. Additionally, web 3.0 and blockchain implementations will either fail to scale or they will be highly centralized. Tokenized web 3.0 approaches introduce unnecessarily highly counter-party economic risk. While decentralized payment systems require both blockchain and Proof-of-Work~\cite{nakamoto2008bitcoin,dembo2020everything} (or smart contracts) for asynchronous distributed consensus under faulty processes, it becomes increasingly clear that other applications including decentralized social media and marketplaces for computational resources work securely and efficiently without blockchain or global consensus. 
In contrast to peer-to-peer networks, web 3.0 and blockchain applications, the NOSTR protocol works efficiently as a social media protocol, because it does not require a majority of nodes of the network to share a common database of the data history. As a consequence, it is also economically efficient since it does not require any form of tokens and it does not hide any counter party risk. In summary, the NOSTR offers a censorship-resistance, open, public network structure, as a well as a digital IDs (DIDs) solution through cryptography and a novel protocol implementation~\cite{liu2023sovereign}. 

  At the core of the NOSTR, there are two main structural components: clients and relays. For social media applications, users run clients to participate in the network and interact/communicate with other users, through the user interface of the client application. This includes, but it is not limited to retrieving posts from other users, posting notes, accessing profiles of other users, re-posts, likes or even sending payments through the lightning network~\cite{poon2016bitcoin}. To guarantee uniqueness of identities, every user is identified by a public key~\cite{nostr-protocol-DID}. Ownership of each identity is guaranteed through a private key that users generate locally~\cite{schnorr}. Every post (or action) is signed by using the private key and every client validates these signatures.

  Users do not communicate directly with each other, but rather through relays and relays interact with users only. Relays passively listen and propagate information. Although relays can not change the content (or action) of users, they can act independently with any other user or relay and refuse to propagate pieces of information. The protocol is robust to such censorship attempts, since anyone including users can run one or multiple relays and users can choose an arbitrary set of relays to interact with. Also anyone can run their own relay. Clients retrieve and publish data from relays of their choice. As an example, assume that the User A "follows" a set S of Users, then the client just queries selected relays for posts from the public key of all users within the set S. That is, on startup a client queries data from all relays it knows for all users it follows (for example, all updates from the last day), then displays that data to the user chronologically. An event (user action, for instance a post or a like) can contain any kind of structured data, and clients choose how to display events (defined as any user generated action) of certain kind, while relays can handle them seamlessly. Examples of different kinds are simple note/post (kind $\#1$), re-post (kind $\#6$), reaction (kind $\#7$)~\cite{nostr-protocol}. The NOSTR protocol offers flexible structure of messages through a customizable structure and different kind of events.

  \section{Related Work}
  The orchestration of middleware systems to distribute computational tasks over wide-ranging networks is a cornerstone in the evolution of distributed computing. A pivotal advancement in this area has been the development of grid computing. Emerging as a potent global cyber-infrastructure, grid computing is engineered to bolster the forthcoming wave of e-Science applications. By integrating large-scale, diverse, and distributed resources, grid computing fosters a cooperative model for the sharing of computational power and data storage on an unprecedented scale. This cooperative paradigm paves the way for unparalleled computational and data management capabilities, enabling the execution of sophisticated scientific, engineering, and academic endeavors previously beyond reach.

Within the grid computing ecosystem, centralized systems such as Pegasus \cite{Deelman2003} and Triana\cite{triana} have emerged as paragons. Pegasus, functioning as a workflow mapping engine, abstracts workflow execution from its underlying infrastructure, facilitating the seamless amalgamation of computational tasks across varied grid environments. This system is heralded for its ability to streamline and optimize complex workflows with remarkable efficiency. Conversely, Triana offers a dynamic platform for the composition of applications, harnessing distributed resources through meticulous integration and orchestration. These frameworks underscore the dynamic and adaptable nature of grid computing, illustrating its capacity to tackle complex challenges through the strategic leverage of distributed resources.

Concurrently, volunteer computing has established itself as an integral component within the broader spectrum of grid systems. The Berkeley Open Infrastructure for Network Computing (BOINC)\cite{boinc_first_paper} exemplifies this model, capturing the essence of volunteer computing by mobilizing the latent processing power of personal computers worldwide for scientific research necessitating extensive computational resources. This model promotes the democratization of scientific inquiry, inviting public engagement while significantly enhancing computational prowess without the need for equivalent infrastructural investment.

The "BOINC: A Platform for Volunteer Computing" \cite{Anderson2019} highlights the enduring success of the BOINC project, demonstrating its sustained operation over the years as a beacon of collaborative scientific endeavor. This analysis brings to the forefront the prohibitive costs and reliability concerns associated with conventional cloud computing platforms, such as Amazon Web Services (AWS), which, despite their widespread adoption, pose financial and operational challenges for extensive, computation-intensive projects. Furthermore, the discourse addresses the intrinsic challenge of maintaining a robust volunteer base — a critical component for the sustenance of volunteer computing frameworks. The fluctuating levels of engagement and the logistical complexities of coordinating such a decentralized workforce underscore the necessity for innovative solutions to bolster volunteer retention and participation.

To address the lack of incentives and participation in volunteer computing, reward-based approaches have been introduced in prior works. For instance, Gridcoin\cite{gridcoin} is a proof-of-stake cryptocurrency that rewards participants for contributing computational power to scientific research projects, notably through the Berkeley Open Infrastructure for Network Computing (BOINC) and Folding@Home. However, a proof-of-stake cryptocurrency may introduce several technical issues, because the payment network requires a global synchronized clock~\cite{dembo2020everything}. Such issues and significant outages become prominent in many non-proof-of-work blockchain systems (for instance~\cite{yakovenko2018solana}).

While blockchain based approaches aim to provide innovative solutions to incentivize contributions in distributed computing, such approaches differ significantly from the NOSTR protocol. NOSTR offers a broader set of capabilities, extending beyond the computational resource sharing model to include secure communications and permissionsless interaction between customers and service providers. In contrast with volunteer computing and proof-of-stake cryptocurrency models, the decentralized protocol NOSTR integrates bitcoin through the lightning network and a dedicated protocol flow for payments. NOSTR solves the user-to-user micro-payment problem, incentivizes honest collaborative computation, enables an open market through instant, permissionless payments and presents a viable solution to the challenge of volunteer maintenance. These, together with the secure and robust digital IDs implementation, make NOSTR the state-of-the-art protocol, capable of supporting a wide range of decentralized applications and services, offering a pathway to enhance viability, sustainability, and effectiveness.


  \section{A Decentralized Marketplace for Distributed Computing}
  As the name of the protocol indicates (Notes and Other Stuff Transmitted by Relays), the type of communication messages and user-to-user interactions are not restricted to simple notes, and they support a variety of alternative applications including direct messaging, file sharing, audio/video streaming, collaborative writing, marketplaces for data processing and more~\cite{nostrapps}. Another example is marketplaces for data processing, where users interact with each other to execute AI algorithms. For the implementation of data processing on the NOSTR protocol, dedicated event kinds have been developed to facilitate distributed data processing (see~\cite{dmvs,vendata}). These implementations have enabled the design of a decentralized AI directory marketplace of the form "Money-In Data-Out". For instance, a customer can provide some input data, such as text or audio, and pay a service provider to receive a short summary of the input data. More specifically, the kinds in range $\#5000-\#5999$~\cite{dmvs} are reserved for text manipulation jobs including text extraction, summarising text, text generation and translation. Image generation corresponds to kind $\#5100$, and video/audio manipulation corresponds to kinds in range $\#5200 -\#5299$.
  
  In this work, we propose an extension of the protocol. Our goal is to design a decentralized marketplace for federated learning and LLMs training. In a nutshell, a customer provides a dataset, model specification and a required payment to a set of service providers. Then the service providers execute the model training, and they return the parameters of the model in order to receive the payment for their work (Money-In AI-Out). In other words, we propose a protocol flow for developing AI-Model Vending Machines (AI-VM), where a customer provides a dataset, a set of specifications, and submits a payment to a set of service providers to acquire a model trained on the input dataset. There are several components of the protocol that allow us to build upon existing protocol rules and design a decentralized marketplace. These include Digital IDs, payments, fast communication though websockets, a flexible protocol structure, also suitable for distributing computing. We proceed by briefly explaining these features. 

\section{Signatures \& Digital IDs}

The NOSTR identity is a dual-key cryptographic system~\cite[Schnorr Signatures]{schnorr, maxwell2019simple}. Each user profile is associated with a private (nsec) and a public key (npub). These keys are generated locally without the need of a trusted party. Cryptographic keys are essential to the protocol's security, ensuring the protection of user identities and messages. The private key is used to sign messages and other actions by the user, verifying that they originate from the authentic identity owner. On the other hand, the public key is employed to authenticate these messages, confirming that they were signed using the corresponding private key. Similarly to social media applications, the same DID, private and public keys can be used for alternative applications, for instance private messaging applications, or marketplaces. This allows for a consistent and secure way to authenticate users DIDs across different applications over the NOSTR protocol.

\paragraph{Nostr Digital Identity and Reputation:} Owning a well-known DID or building a solid reputation over the NOSTR offers the benefits of recognizability and trust from other users, similarly to all other online media platforms. Notably, this identity and its accompanying reputation can be seamlessly transferred between different applications. In the context of a marketplace, a good reputation can help service providers maximize their profit. A high reputation indicates a consistent service, which in turn makes more customers trust the service provider, ensuring a steady stream of business.  Alternatively, bad reputation may assist customers to avoid incompetent service providers. As a matter of fact, a reputation system for a decentralized and open marketplace for distributing computing will be valuable. However, this remains outside the scope of this work, but it is an interesting implementation possibility for future extensions of the protocol.

\section{Protocol Design for Distributed Computing and AI Directories}

The NOSTR protocol supports a preliminary version of AI directories for data processing in the form of an open and decentralized marketplace~\cite{dmvs,vendata}, known as Data Vending Machines. On one side, there are users (customers) who provide some data that should be processed (image, video or text) and they request jobs that can be done with some AI tools, for instance image generation from text, a transcript of a video, a summary of an article. On the other side, there are AI agents (service providers) that offer their services to users, and in exchange for a payment, they deliver the requested results in the form of processed data. This system works in the form of Money-In Data-Out; a marketplace for data processing/manipulation by applying AI tools. This design sets the foundations of distributed data processing over the NOSTR protocol and offers a unique protocol structure and flow for additional development. 

Our goal is to utilize the existing architecture and to go one step further by extending components of the protocol if necessary, in order to design a decentralized marketplace for Federated Learning (FL) and Large Language Models (LLM) training. In what we propose later (see Section~\ref{Section Market}), users (customers) provide an input dataset and certain specifications and request a sequence of jobs that aim to train a deep neural network or an LLM model. Then service providers receive the data, model specifications and the requested task (train through regular FL or LLM training (e.g., via the DiLoCo algorithm)~\cite{douillard2023diloco}), they train the model through multiple rounds and jobs requests, and they return the model parameters. In exchange for their computational resources, the service providers receive recurrent payments from the customer upon delivering valid results. The main difference between our design and existing approaches is the concept of delivering a trained model instead of processed data; a set of service providers are required to optimize the same model in parallel and deliver valid results through multiple rounds of computation. For that purpose we extend the structure of the protocol by introducing new event kinds, a new protocol flow and we deploy the classical FL approach, as well as the DiLoCo algorithm~\cite{douillard2023diloco} for LLM training. Next, we delve into the existing protocol structure for distributing computing over the NOSTR network. We also emphasize the modifications we introduce to support the Money-IN AI-Out marketplace design.

\subsection{ Distributed Computing on the NOSTR Protocol}
Herein, we discuss the existing dedicated protocol messages and protocol flow for distributed computing. We mainly focus on five components that we will utilize later with slight extensions or variations, these include job request events, job result events, job feedback events, events for service provider discoverability, and job chaining rules. These type of events consist the core of the communication messages between customer and service providers. We refer the reader to the NOSTR Implementation Possibility - 90 (NIP-90~\cite{nostr-protocol-DVM}) for further details. 

\paragraph{Basic Protocol Flow for Data Vending Machines (DVMs)~\cite{nostr-protocol-DVM}} 

\begin{itemize}
\setlength\itemsep{0em}
    \item Customer publishes a job request (e.g. kind:5000, speech-to-text).
    \item Service Providers may submit job feedback events (kind:7000, e.g. payment-required, processing, error, etc.).
    \item Upon completion, the service provider publishes the result of the job with a job-result event (kind:6000).
    \item At any point, if a payment is pending as instructed by the service provider, the user can pay the bolt11 invoice~\cite{antonopoulos2021mastering} included in the corresponding job result event.
\end{itemize}

The protocol is designed to be intentionally flexible. For instance, a customer may choose a set of service providers and restrict or not restrict the job request to be completed by this specific set of service providers. Also, a service provider may not start a job until they receive a form of payment, or their response can depend on the other participants reputation based on their public key. The flexibility of the protocol flow allows us to extend parts of it, and implement additional communication steps for the design of on demand LLMs training over a decentralized marketplace. Next we proceed by presenting the existing events as appear in the protocol flow for DVMs.

\paragraph{Job Request (DVMs)} A Job Request for Data Vending Mchines is an event, published by a customer through a set of relays. This event signals that a customer is interested in receiving the result of some kind of computation by an AI tool. The reserved range for job request events is $\#5000 -\#5999$. The field with label "content" is empty, while, all the required information should be provided in the field with label "tags". The field "tags" contains all the required information from an AI-agent in order to process the data.
\begin{lstlisting}[escapeinside={(*}{*)}, caption={Job Request Event with kind in range 5000 -5999},captionpos=b, label ={JRLabel}]
{
  "kind": 5xxx // kind in range 5000 - 5999,
  "content": "",
  "tags": (*$\big[$*)
     (*$\big[ \text{"i", " <data> ", "<input-type>, "<relay>", "<marker>"} \big]$*),
     (*$ \big[ \text{"output", "<mime-type>"} \big]$*),
     (*$ \big[ \text{"relays", "wss://..."} \big]$*),
     (*$ \big[ \text{"bid", "<msat-amount>"}\big]$*),
     (*$ \big[\text{"t", "bitcoin"} \big]$*)
  (*$\big]$*)
}
\end{lstlisting}


\noindent For completeness, we provide additional explanation regarding each tag of the events. We first present the existing events structure. Then we introduce any necessary changes or additions that we will consider in the protocol flow and in the algorithm. For the Job Request event (Event Type \ref{JRLabel}) there are the following tags:  
\begin{outline}
    \1 <i>: All the required input data for the job appear here. Specifically, the inputs of the AI algorithm.
    \vspace{-0.15cm}
      \2  <data>: The argument for the input.
       \2 <input-type>: The type of the argument input, it must be one of the following:
        \3    url: A URL to be fetched of the data that should be processed.
         \3   event: A Nostr event ID.
          \3  job: The output of a previous job with the specified event ID. The determination of which output to build upon is up to the service provider to decide (e.g. waiting for a signaling from the customer, waiting for a payment, etc.)
          \3  text: <data> is the value of the input, no resolution is needed
          \vspace{-0.15cm}
       \1 <relay>: The relay where the event/job was published
       \vspace{-0.15cm}
       \1 <marker>: An optional field indicating how this input should be used within the context of the job
       \vspace{-0.15cm}
    \1 <output>: Expected output format. 
    \vspace{-0.15cm}
    \1 <param>: Optional parameters for the job. 
    Different job request kind defines this more precisely. (e.g. [ "task", "run option", "initial state" ]). 
    \vspace{-0.15cm}
    \1 <bid>: Customer MAY specify a maximum amount (in millisats) they are willing to pay
    \vspace{-0.15cm}
    \1 <relays>: List of relays where Service Providers SHOULD publish responses to.
    \vspace{-0.15cm}
    \1 p: Service providers the customer is interested in. Other service providers might still choose to process the job.

\end{outline}

\paragraph{Job Request (AI VMs)} The existing job request events support a large number of different jobs, dedicated to data processing applications. Such job request have a certain structure (see Event Type \ref{JRLabel}). For clarity and the efficacy of our approach, we propose to introduce the kinds 8000 - 8999 as Job Request events for federated learning, LLM training and distributed optimization. Although each kind can correspond to a different algorithm or variation of an algorithm, specific details can be considered by introducing additional fields in the tags. As an example, notice that the predefined structure of Event Type \ref{JRLabel}, requires a single input data type as we discussed earlier, while additional information may be included in parameters field (<param>).

\begin{lstlisting}[escapeinside={(*}{*)}, caption={Job Request Event for training AI models, kind in range 8000 -8999},captionpos=b, label ={JRLabel_AI}]
{
  "kind": 8xxx // kind in range 8000 - 8999,
  "content": "",
  "tags": (*$\big[$*)
     (*$\big[ \text{"i", "<data>", "<input-type>, "<relay>", "<marker>"} \big]$*),
     (*$ \big[ \text{"output", "<model-parameters-type>",} \big]$*),
     (*$ \big[ \text{"relays", "wss://..."} \big]$*),
     (*$ \big[ \text{"bid", "<msat-amount>"}\big]$*),
     (*$ \big[\text{"t", "bitcoin"} \big]$*)
     (*$ \big[\text{"p", "<service-provider(s)-public-key(s)>"} \big]$*)
     (*$ \big[\text{"param", \color{red}"task", "Inner-or-Outer"}\big]$*),
     (*$ \big[\text{"param", \color{red}"run option", "Fevavg-or-DiLoCo"}\big]$*)
     (*$ \big[\text{"param", \color{red}"data\_set", "<URL>"}\big]$*),
     (*$ \big[\text{"param", \color{red}"initial/current-model-state", "<raw-data>"}\big]$*)
     (*$ \big[\text{"param", \color{red}"model", "LLaMA-2"}\big]$*),
     (*$ \big[\text{"param", \color{red}"source\_code", "<URL>"}\big]$*),
     (*$ \big[\text{"param", \color{red}"expected\_execution\_time", "<time>"}\big]$*),
     (*$ \big[\text{"param", \color{red}"recommended\_hardware\_specification", "<text>"}\big]$*),
     (*$ \big[\text{"param", \color{red}"validation\_rules\_for\_the\_output", "<URL>"}\big]$*)
     (*$ \big[\text{"param", \color{red}"timeout-specification", "max-time>"}\big]$*)
  (*$\big]$*)
}
\end{lstlisting}
Similarly, for the purpose of training an LLM, the customer needs to provide a training dataset through a URL as input data, and additionally a set of parameters related to the training task, for instance, initial model parameters, a URL to the source code of the training or optimization method, model specification (example "LLaMA-2"), rules for validation of the output. To maintain the existing protocol flow as much as possible, we consider the input data as a URL to the training dataset for the initial job request, or the output of a previous job with a specified job event ID to indicate a job chaining and multiple rounds of training. In both cases (initial or chain job request), we consider all the required information for algorithm execution in the parameters fields for consistency. Below some pieces of that information follow (Event Type \ref{JRLabel_AI}): 

\begin{itemize}
\setlength\itemsep{0em}
    \item a source for the code: While the service provider may have its own code, there should be a reference to a standard approach that guarantees consistency
    \item a source for the data: The customer splits and sends an encrypted URL to the service providers. The customer decides how to split the data. The way of splitting the data may be arbitrary.  
    \item job details: expected execution time or required hardware, rules related to timeout conditions.
    \item Declaration of validation process. This is how the customer will decide if the output is accurate or not. 
\end{itemize}


Through the optimization process, we introduce a new job request at each round. We also consider job chaining to speed up the process and most likely to execute the training by assigning the jobs to the same service providers, as long as they provide valid outputs. A running job will be reassigned to a new service provider if a previous service provider goes offline, or fails to provide valid result. Finally, the customer can encrypt all parameters so that only the selected service provider(s) in tag "p" can decrypt them. We proceed by briefly discussing the Job Chaining rules, for a detailed description we refer the reader to~\cite[(NIP-90)]{nostr-protocol-DVM}.

\paragraph{Job Chaining} A customer has the option to request multiple rounds of jobs to be processed as a chain, where the output of a job is the input of another job. In the context of data processing and DVMs this can be a sequence of processing as a podcast transcription by one service provider and then extracting a summary of the transcription by another. One way to implement this is by specifying an event ID of a previous job as input, using the job type field. In general, service providers may begin processing a subsequent job the moment they see the prior job result, but they will likely wait to receive a (partial) payment first.

In the case of federated learning and LLM training such options still exist; however, we propose certain protocol rules to minimize the risk introduced by bad actors and malicious users. To do this, we consider it necessary for the customer to validate the progress made by each service provider at every single round. After the validation is completed, each service provider who produced successful results receives a partial payment for the work done in the corresponding round. Then the customer updates the current state of the model and submits a follow-up job request for the next round (and optionally a partial payment to the service providers upfront). Upon completion of the optimization task at each round, the service providers communicate with the customer through a Job Result event.   



\paragraph{Job Result (kind: 6000-6999)} Service providers communicate with the customer through Job Result events, providing the (optionally encrypted) output of the job. In the case of AI training, the output corresponds to an optimized version of the model parameters. The format of the Job Result event appears in the Event Type \ref{JRE} above. The tags of the Job Result event include the original job request event id (stringified-JSON), the public key of the customer, amount that the Service Provider is requesting to be paid or an lightning invoice (bolt11~\cite{antonopoulos2021mastering,bolt11_git}), and the optimized model parameter as an encrypted output. Finally, an "i" tag provides additional information related to the initial job request or for validation of the output by the customer.
\begin{lstlisting}[escapeinside={(*}{*)}, caption={Job Result Event}, captionpos=b, label={JRE}]
{
  "pubkey": "<service-procider pubkey>",
  "content": "<enctypted payload>",
  "kind": 6xxx // kind in range 6000 - 6999,
  "tags": (*$\big[$*)
     (*$\big[ \text{"request", " <job-request> "} \big]$*),
     (*$ \big[ \text{"e", "<job-request-id>", "<relay-hint>"} \big]$*),
     (*$ \big[ \text{"p", "<customers's-pubkey>"} \big]$*),
     (*$ \big[ \text{"amount", "requested-payment-amount", "<optional-bolt11>"}\big]$*),
     (*$ \big[ \text{"i", "additional-information-for-validation", "<information>"}\big]$*)
     (*$ \big[\text{"output", "encrypted: \color{red} model parameters (and loss for validation purposes)"} \big]$*)
  (*$\big],$*)
  ...
}
\end{lstlisting}



\paragraph{Job Feedback (kind:7000)} During the model training process, service providers can communicate feedback for an ongoing job through the Job Feedback event. Customers may demand job feedback to avoid timeouts and ensure that a service provider has achieved some partial progress. Information about partial progress can be optionally included by adding fields to the tags in Event Type \ref{Job_Feedback_Event}.
\begin{lstlisting}[escapeinside={(*}{*)}, caption={Job Feedback Event},captionpos=b, label ={Job_Feedback_Event}]
{
  "kind": 7000,
  "content": "<empty-or-payload>",
  "tags": (*$\big[$*)
     (*$\big[ \text{"status", " <status> ", "<extra-info>"} \big]$*),
     (*$ \big[ \text{"amount", "<requested-payment-amount>", "<bolt11>"} \big]$*),
     (*$ \big[ \text{"e", "job-request-id", "<relay-hint>"} \big]$*),
     (*$ \big[ \text{"p", "<customer's-pubkey>"}\big]$*),
  (*$\big],$*)
  ...
}
\end{lstlisting}

The predefined format of the Job Feedback event includes the following tags~\cite{nostr-protocol-DVM}:
\begin{outline}
\1    <content>: It may be empty, a final job result, or a partial result  (for a job in progress). This field will have an encrypted payload with p tag as key.
\vspace{-0.15cm}
  \1  <amount>: Requested payment amount.
\vspace{-0.15cm}
    \1 <status>: The service provider publishes the current status of the requested job. For instance:
    \vspace{-0.15cm}
        \2 payment-required: Service Provider requires payment before continuing
        \2 processing: Service Provider is processing the job
         \2 error: Service Provider was unable to process the job
         \2 success: Service Provider successfully processed the job
         \2 partial: Service Provider partially processed the job. The content filed might include a sample of the partial results.
\end{outline}


\paragraph{Discoverability} Service Providers can announce their availability or to advertise their support for specific job kinds by publishing discoverability events~\cite[(NIP-89)]{nip89}. The format structure of the discoverability event appears in the Event Type \ref{Discover_Event}. By including an "i" tag, service providers may provide additional information related to their available computational capabilities, hardware specifications, or execution time limits. 
\lstset{linewidth=15cm}
\begin{lstlisting}[escapeinside={(*}{*)}, caption={Discoverability Event},captionpos=b, label ={Discover_Event}]
{
  "kind": 31990,
  "pubkey": "<pubkey>",
  "content": (* \big[ *)
    \"name\": (* $ \text{\ "Federated Learning AI-VM" \ } $ *),
    \"about\":  (* "I'm a AI-VM for federated learning.*)"
  (* \big] *) ,
  "tags": (*$\big[$*)
     (*$\big[ \text{"k", "8000"} \big],  \text{ // e.g. optimization methods, federated learning}$*) 
     (*$ \big[\text{"t", "bitcoin"} \big]$*)
     (*$ \big[\text{"i", "specifications", "hardware", "maximum-execution-time", "model-dimensions-range"} \big]$*)
  (*$\big],$*)
  ...
}
\end{lstlisting}

\section{Payments}


The NOSTR protocol includes integrated payment functionality; requests for payment, invoice notifications, receipts for payments, as well as payments to certain events or directly to a user or a group of users. The payments take place through the lightning network~\cite{poon2016bitcoin}
. For any form of application, including social media, Internet-of-Things applications~\cite{o2022money}, and marketplaces~\cite{robert2020enhanced}, the lightning network enables bitcoin payments with instant settlement and virtually zero transaction fees. These properties, together with in-channel messages, are ideal for reliable handshaking mechanisms, anti-spam and on the fly payments between multiple job rounds among devices. 

To further explain how the payments blend in with the rest of the protocol, we proceed with an example. In the case of federated learning, the aggregator itself may act as a customer who pays service providers to optimize a model on different subsets of a training dataset through multiple job rounds. Initially, the customer (aggregator) identifies possible service providers through the discoverability event (kind 31990, Event Type \ref{Discover_Event}), selects a number of them and submits a Job Request event (kind $\#8000$, event type~\ref{JRLabel_AI}). The service providers respond with a Job Feedback event (kind $\#7000$, event type~\ref{Job_Feedback_Event}) with status payment-required, to announce that they are available to complete the job. Then the customer proceeds by submitting a small amount (of the total) as a payment and the service provider starts the job. Upon delivery of the final or partial output result (Job Result \ref{JRE} \& Job Feedback \ref{Job_Feedback_Event} events), the customer proceeds with a partial or total payment. This process repeats for each optimization round and for all service providers (assuming that they provide valid results). The ability to break down payments into partial amounts and pay upon delivery of a valid result significantly reduces the possibility that payments will be executed without completing the requested task. This strategy, together with a reputation system, can make the marketplace robust against adversarial or spam attacks to a large extent. However, an in-depth implementation of such sophisticated strategies remains out of the scope of this work. We discuss strategies for validation of the delivered output in Section \ref{Section: Cust}.

\subsection{Payments Implementation}

In this section, we discuss the payment system which is integrated into the NOSTR protocol. We briefly explain three components of the payment system; Payment Request, Payment Receipts and Validation of Receipts. For detailed documentation of the payment protocol flow, we refer the reader to implementation 57~\cite{nip57,nostrzaps}. 
\paragraph{Payment Request}
A payment (known as zap) request is an event of kind 9734 that is not published to relays, but is instead sent to a lnurl pay callback url~\cite{lnurlpay} of the recipient. The field content may be an optional message to send along with the payment. The payment request event must include the following tags:
\begin{itemize}
  \setlength\itemsep{0em}
    \item Relays: A list of relays where the recipient's wallet should publish its zap receipt. 
    \item Amount: The payment amount in millisats that the sender intends to pay, formatted as a string. 
    \item Lnurl: The lnurl pay url of the recipient, encoded using bech32~\cite{antonopoulos2021mastering} with the prefix lnurl. 
    \item p: The hex-encoded pubkey of the recipient.
    \item e: An optional hex-encoded event ID. Implementations must include this when paying for an event rather than directly to a public key.
\end{itemize}

%

\begin{lstlisting}[escapeinside={(*}{*)}, caption={Payment Request Event},captionpos=b, label ={zap_request_event}]
{
  "kind": 9734,
  "content": "Zap!",
  "tags": (*$\big[$*)
    (*$\big[$*)"relays", "<wss://relay-domain>", "wss://anotherrelay.com"(*$\big]$*),
    (*$\big[$*)"amount", "<msats-amount>"(*$\big]$*),
    (*$\big[$*)"lnurl", "<a-static-lightning-invoice>"(*$\big]$*),
    (*$\big[$*)"p", "<hex-encoded-pubkey-of-the-recipient>"(*$\big]$*),
    (*$\big[$*)"e", "<optional-hex-encoded-event-ID>"(*$\big]$*)
  (*$\big]$*),
  "pubkey": "<NPUB>",
  "created_at": <timestamp>,
  "id": "<event-ID>",
  "sig": "<signature>"
}
\end{lstlisting}

\paragraph{Payment Receipt}

A payment receipt is created by a lightning node when an invoice generated by a payment request is paid. Payment receipts are only created when the invoice description (which is committed to the description hash) contains a payment request note. When receiving a payment, a series of steps is executed, and a NOSTR event of kind 9735 (shown below as Event Type \ref{zap_receipt_event}) should be published to the relays declared in the payment request. We refer the reader to the detailed documentation~\cite{nip57} for a complete description of the Payment Receipt event structure and the full protocol flow. 

\begin{lstlisting}[escapeinside={(*}{*)}, caption={Payment Receipt Event},captionpos=b, label ={zap_receipt_event}]
{
    "id": "<event-ID>",
    "pubkey": "<NPUB>",
    "created_at": <paid-at-date>,
    "kind": 9735,
    "tags": (*$\big[$*)
      (*$\big[$*)"p", "<payment-recipient)>"(*$\big]$*),
      (*$\big[$*)"P", "<optional-P-tag-from-the-pubkey-of-the-payment-request>"(*$\big]$*),
      (*$\big[$*)"e", "<optional-tag-same-as-payment-request-e-tag>"(*$\big]$*),
      (*$\big[$*)"bolt11", "<lightning-invoice>"(*$\big]$*),
      (*$\big[$*)"description", "<SHA256(description)-from-invoice>"(*$\big]$*),
      (*$\big[$*)"preimage", "<payment-preimage>"(*$\big]$*)
    ],
    "content": "",
  }
\end{lstlisting}

\paragraph{Receipt Validation}

The service provider can retrieve payment receipts using a NIP-$01$ filter~\cite{NIP-01} and validate each receipt to verify that a customer fetched their invoice and their payment claim for a certain job. Payment validation follows these steps:
\begin{itemize}
\setlength\itemsep{0em}
    \item The public key in the payment receipt event must match the public key of the service provider.
    \item The invoice amount contained in the bolt11 tag of the payment receipt must equal the amount in the amount tag of the payment request.
    \item The lnurl tag of the payment request should be equal to the lnurl of the recipient.
\end{itemize}

\noindent We highlight that a receipt is not proof of payment; it only proves that some NOSTR user fetched an invoice. As a final validation step, the recipient must verify the payment on their Lightning node or payment processor.

\section{Money-In AI-Out: Marketplace for Federated Learning and LLMs}\label{Section Market}
Herein, we introduce the protocol flow and algorithmic design for federated learning and LLMs training over the NOSTR protocol. By leveraging the existing structure of the protocol for data processing, and by introducing additional components when it is necessary, we construct an open and decentralized marketplace for training AI models. In this market, a customer provides a dataset and model specifications. The service providers (AI Vending Machines) receive a payment, and return an AI model trained on this dataset. Customers and service providers communicate through relays, which can be public or private (and self hosted). Additionally, the communication can use encryption to guarantee privacy for the dataset or the computational output. We proceed by explaining further the protocol flow and then we present a detailed algorithmic implementation for customers and service providers.

We propose a protocol flow involving multiple job rounds between customers and service providers. Through multiple job rounds, we ensure valid progress in the optimization process of each service provider individually. This enables on-the-fly payments after validation of the output at each round. If a service provider delivers an invalid output or delays the job beyond a certain time limit, the customer reassigns the job to another service provider without paying the invalid (or delayed) job output.

In this work, we consider two training methods: the classical federated learning~\cite[FEDAVG]{mcmahan2017communication} and the DiLoCo algorithm~\cite{douillard2023diloco} for training LLMs. Both of these have an inner and outer optimization phases. This allows us to present our approach in parallel. Specifically, we define the job round as the successful sequence of the following events and actions. For the first part (inner optimization): job request, inner optimization computation executed by the service providers, retrieval of outputs by the customer through job result events, validation of the outputs. For the second part (outer optimization): use the latest inner optimization outputs as inputs, execute the outer optimization (either through a job request, or executed locally by the customer), complete any pending payments and update the model parameters for the next round (if any). Upon validation of the outputs, a (partial) payment is submitted by the customer to each service provider. To explain this further we provide the sequence of the steps below.
\paragraph{Protocol Flow for Each Job Round.} At every job round:
\vspace{-0.2cm}
\begin{outline}
    \1 Customer (Algorithm \ref{alg:customer}) publishes a job request (e.g. kind 8000: Federated Learning, inner optimization). The job request may include a list of chosen service providers, and it must include all the required information: the current state of the model (for instance for the first round initial value of model parameters), a link (source) to the data, a list of relays where the service providers should publish their job results job feedback at, as well as all the required parameters (see Event Type \ref{JRLabel_AI}).
\vspace{-0.15cm}
    \1 Service Providers (Algorithm \ref{alg:service_provider}) must submit job feedback events (kind 7000, e.g. partial payment required) before they start the job with STATUS = payment\_required, if they wish a (partial) payment before starting computation (Event Type \ref{Job_Feedback_Event}). This requests a partial payment and signals that the service provider is ready to start upon receiving it.
\vspace{-0.15cm}
    \1 Customer submits an initial payment to service providers (responsible for inner optimization).
    \vspace{-0.2cm}
    \1 Upon receiving the partial payment the service providers start executing the inner optimization job (Algorithm \ref{alg:inner}). 
    \2 Within a fixed time period, service providers publish Job Feedback events (Event Type \ref{Job_Feedback_Event}) with partial/sample results.
\vspace{-0.15cm}
    \2 If results delivery exceeds a predefined time limit, the customer stops the procedure and seeks other service providers (responsible for inner optimization).
    \vspace{-0.2cm}
    \1 Upon completion, each service provider publishes the output of the job through a Job Result event (kind 6000, Event Type \ref{JRE}).
\vspace{-0.15cm}
    \1 Customer validates the results/ouputs (Algorithm \ref{alg:validation})
\vspace{-0.15cm}
    \2 If the output is valid, then the customer completes payments to all service providers (responsible for inner optimization) who returned valid outputs.
\vspace{-0.15cm}
    \2 If the result is not valid, the customer reassigns the job to a different service provider (Algorithm \ref{alg:reassign}).
\vspace{-0.15cm}
    \1 Upon receiving successful outputs for all inner optimization jobs, the customer proceeds to the outer optimization (Algorithm \ref{alg:outer}).
\vspace{-0.15cm}
    \2 If the customer selects to execute the outer optimization locally, then the customer executes the outer optimization (Algorithm \ref{alg:outer}) with input the output of inner optimization procedures.
\vspace{-0.15cm}
    \2 Otherwise, if the customer chooses to assign the task to a service provider for the outer optimization, the customer follows the steps for reassigning the job from SELF to a service provider (Algorithm~\ref{alg:reassign}), retrieving and validating the output, and completing any payments identically to the inner optimization job handling as described above.

    



\vspace{-0.15cm}
    \1 At the end of each round, the customer updates the model parameters to use them as input for the next round.
\vspace{-0.15cm}
    \2 If the current round is the terminal round, then the customer completes any pending payments and ends the procedure outputting the final result.
\vspace{-0.15cm}
    \1 At any point, if there is an amount pending payment as instructed by the service provider, the customer can pay the Lightning-invoice~\cite[bolt11]{antonopoulos2021mastering,bolt11_git} (see lnurl Event Type \ref{zap_request_event}).
\end{outline}
We present an implementation of the protocol flow in Section \ref{Section: Cust} and Section \ref{Section: SP}.
\subsection{Customer}\label{Section: Cust}
A complete implementation of the customer routine appears in Algorithm~\ref{alg:customer}. This includes the steps followed and the events published by the customer. First, the customer discovers and requests a set of service providers to run the inner optimization part of the job, as explained in the protocol flow above. Upon successful delivery of outputs, the customer proceeds to the outer optimization. The outer optimization for both FedAvg and DiLoCo is a single or double update of local variables respectively. However we consider the option for the customer to also assign this computation step to a service provider. This can be useful in practice if a customer prefers to implement an outer optimizer with higher computational complexity (for instance, multiple iterations) but is unable to perform such computations locally. We now explain three parts of the main customer routine in detail: (1) how the customer handles disconnections and failures to respond (Algorithm \ref{alg:customer}), (2) how the customer validates the output results provided by a service provider at job completion (Algorithm \ref{alg:validation}), and (3) how the customer reassigns a job to a different service provider if needed (Algorithm \ref{alg:reassign}).

\paragraph{Handling Disconnections and Failures to Respond.} The customer may consider a timeout period. If a service provider does not provide the output within a certain time window (from receiving the initial payment), the customer reassigns the job to a different service provider (Algorithm~\ref{alg:reassign}). As an example of the timeout implementation, see Algorithm~\ref{alg:customer} (lines 13-14). All information regarding the timeout rules should be included in the Job Request event (Event Type~\ref{JRLabel_AI}).

The definition of failure to respond may vary from one implementation to another. As an example, the customer may choose to receive a valid output within a certain time period. Alternatively, the timeout decision may depend on periodic responses from the service provider through Job Feedback events (Event Type \ref{Job_Feedback_Event}), which notify the customer about the status of the job and provide partial computation results.

\paragraph{Validation of the Output (Algorithm \ref{alg:validation}).} 
 Upon receiving a Job Result event (Event Type \ref{JRE}), the customer proceeds to verify the accuracy of the output. In the context of training AI models with FEDAVG and DiLoCo, we consider two conditions that test the loss decay on a validation dataset. Specifically, Algorithm \ref{alg:validation} requires as input the validation dataset, the current state of the model (model parameters), and a threshold for the two following policies. As a first option, we compare the loss decay relative to other service providers (see Accuracy Against Service Providers, Algorithm \ref{alg:validation}). This validation test has been introduced in~\cite[Accuracy Checking]{pmlr-v97-bhagoji19a}. As a second option, we consider a test that checks whether a moving average of the loss decays at a certain rate (Algorithm \ref{alg:validation}). Alternative accuracy tests and approaches for identifying malicious service providers appear in prior works~\cite{demartis2022adversarial,pmlr-v97-bhagoji19a,rathee2023elsa,awan2021contra}. Such implementations could be integrated with our design but remain outside the scope of this work.
\begin{algorithm}
\caption{Customer}\label{alg:customer}
\begin{algorithmic}[1]
\Require \cblue{C\_NPUB}: Customer's NOSTR public key, \cblue{DATA\_SOURCE}: link to the data, \cblue{NUM\_PR}: \# of providers, \cblue{NUM\_JOBS} : Total \# of jobs for each provider, \cblue{RELAYS}: a set of relays 
\Ensure Connection through websockets to all \colorvar{\textit{relays}} in \cblue{RELAYS} \Comment{\color{brown} Where events are published at\color{black}}
\State \textbf{Search Routine}: Search for \cblue{NUM\_PR} \cblue{AVAILABLE} service providers. \Comment{\color{brown} Event Type \ref{Discover_Event}\color{black}} \\ \textbf{Search Routine returns}: \cblue{PROVIDERS} with \cblue{NUM\_PR} NPUBs entries

\State Split the data into \cblue{NUM\_PR} parts and generate the list \cblue{DATA\_SOURCES} with \cblue{NUM\_PR} sources for each data segment \Comment{\color{brown} \cblue{DATA\_SOURCES}\color{black}$(i)$ \color{brown} is the data source of the $i_\text{th}$ provider \color{black}} 
\State Initialize the optimization parameters and model: \colorvar{\textit{model\_parameters}}, \colorvar{\textit{model}}, ...
\State Publish \colorkind{Job Requests} \colorkind{KIND: 8000} with all the required information \Comment{\color{brown} Event Type \ref{JRLabel_AI} \color{black}}
\State Wait for \colorkind{Job Feedback} \colorkind{KIND: 7000} and payment-request from service providers \Comment{\color{brown}Event Type \ref{Job_Feedback_Event} \color{black}}
\State Submit an initialization payment to \cblue{PROVIDERS} 

\For{\colorvar{\textit{job}} in (1,\cblue{NUM\_JOBS})} 

    \State Initialize \colorvar{done} $\leftarrow$ FALSE
    \While{\colorvar{done} $\neq$ TRUE }
    \State Wait some time
    \State Fetch \colorkind{Job Feedback} events \colorkind{(KIND:7000)} from each provider in \cblue{PROVIDERS}
    \If{timeout occurs} 
    \State Call Reassign (Algorithm \ref{alg:reassign}) to replace all the delayed service providers, update \cblue{PROVIDERS} 
    \EndIf
    \If{\cblue{JOB\_STATUS} = \colorvar{success} for all providers in \cblue{PROVIDERS}}
    \colorvar{done} = TRUE
    \EndIf
    \EndWhile
    \State Fetch \colorkind{Job Results} (\colorkind{KIND: 6000}) and \colorvar{"outputs"} from all providers in \cblue{PROVIDERS} \Comment{\color{brown}Event Type \ref{JRE} \color{black}}
    \For{\colorvar{\textit{service\_provider}} in \cblue{PROVIDERS} call the Validation routine (Algorithm \ref{alg:validation}) and}
        \While {Validation\_Test = \cblue{FAIL} for the \colorvar{\textit{service\_provider}}} 
        \State  Reassign the job of \colorvar{\textit{service\_provider}}; call Reassign (Algorithm \ref{alg:reassign}), update \cblue{PROVIDERS}
        \EndWhile
        \State Submit (partial) payments to \colorvar{\textit{service\_provider}} and publish \colorkind{Payment Receipt (KIND: 9735)}
    \EndFor
    \If {\cblue{OUTER\_OPTIMIZER} $=$ \cblue{SELF}}    
        \State Run the Outer\_Optimization routine (Algorithm \ref{alg:outer}) with input the current updates \colorvar{"outputs"} 
    \Else 
        \State Reassign the Outer Optimization from \cblue{SELF} to a service provider (Algorithm \ref{alg:reassign})
        \State Submit (partial) payment to \cblue{OUTER\_OPTIMIZER} and publish \colorkind{Payment Receipt (KIND: 9735)}
    \EndIf
    
    \State Update \textit{\colorvar{model\_parameters}} using the "output" of the OUTER\_OPTIMIZER 
    \If {stopping condition is satisfied } 
            \State break
    \ElsIf{\colorvar{\textit{job}} $\leq $ \cblue{NUM\_JOBS}}
        \State Publish \colorkind{Job Requests} \colorkind{KIND: 8000} with all the required information \Comment{\color{brown} Event Type \ref{JRLabel_AI} \color{black}}
    \EndIf

\EndFor
\State Finalize payments to every \colorvar{\textit{service\_provider} in \cblue{PROVIDERS}} and publish receipts \Comment{\color{brown} Event Type \ref{zap_receipt_event} \color{black}}
\\ \Return \colorvar{\textit{model\_parameters}}
\end{algorithmic}
\end{algorithm}
\paragraph{Reassigning the Job to a Different Service Provider (Algorithm~\ref{alg:reassign}).} Under a timeout or a failed validation instance, the customer reassigns the job to a different service provider. We provide an example of such implementation in Algorithm \ref{alg:reassign}. Further, under perpetual timeouts and invalid output detection, the customer continues to reassign the job, until a service provider delivers a valid output. Then the routine returns the public key of new service provider and the customer updates the list of the service providers under consideration. 



\begin{algorithm}
\caption{Reassign the Job of a Service Provider to a Different Service Provider}\label{alg:reassign}
\begin{algorithmic}[1]
\Require \cblue{SP\_NPUB}: public key of the service provider that will be replaced, <computation information \& time>, \cblue{RUN\_OPTION}: <FEDAVG or DiLoCo>, \cblue{DATA:} \cblue{DATA\_SOURCES} , \cblue{PROVIDERS}, \cblue{TASK}: <INNER or OUTER>
\Ensure Connection through websockets to all \colorvar{\textit{relays}} in \cblue{RELAYS} \Comment{\color{brown} Where events are published at\color{black}}

\State \textbf{Search Routine}: Search for an \cblue{AVAILABLE} service provider for \colorkind{KIND: 8000}. \Comment{\color{brown} Event Type \ref{Discover_Event}\color{black}} \\ \textbf{Search Routine returns}: A single public key: \cblue{SP\_NPUB\_NEW}  

\State Publish a \colorkind{Job Request} \colorkind{KIND: 8000}: identical to the last job request for the service provider \cblue{SP\_NPUB}, but with updated field "p" set to \cblue{SP\_NPUB\_NEW} \Comment{\color{brown} Event Type \ref{JRLabel_AI}\color{black}}
\State Wait for \colorkind{Job Feedback} \colorkind{KIND: 7000} and payment-request from service provider \cblue{SP\_NPUB\_NEW}
\State Submit an initialization payment to \cblue{SP\_NPUB\_NEW}

\While{\cblue{JOB\_STATUS} $\neq$ \cblue{success} for \cblue{SP\_NPUB\_NEW}} 
    \State Wait some time
    \State Fetch \colorkind{Job Feedback} events \colorkind{(KIND:7000)} from service provider \cblue{SP\_NPUB\_NEW} \Comment{\color{brown} Event Type \ref{Job_Feedback_Event} \color{black}}
    \If{timeout occurs} break and reassign again to replace \cblue{SP\_NPUB\_NEW}
    \EndIf
    \If{\cblue{JOB\_STATUS} = \colorvar{success} for \cblue{SP\_NPUB\_NEW}} 
    \State Fetch \colorkind{Job Result} (\colorkind{KIND: 6000}) and \colorvar{"output"} from \cblue{SP\_NPUB\_NEW} \Comment{\color{brown} Event Type \ref{JRE} \color{black}}
    \State Call the Validation Routine
        \If{Validation\_Test = \cblue{PASS}} 
        \State \Return \cblue{SP\_NPUB\_NEW}, "output"
        \ElsIf{Validation\_Test = \cblue{FAIL}} break and
         reassign again to replace \cblue{SP\_NPUB\_NEW}
        \EndIf
    \EndIf
\EndWhile

\end{algorithmic}
\end{algorithm}

\begin{algorithm}
\caption{Validation of the Service Provider Output}\label{alg:validation}
\begin{algorithmic}[1]
\Require \cblue{SP\_NPUB}: service provider public key, \cblue{RUN\_OPTION}: <FEDAVG or DiLoCo>, validation dataset: $\mathcal{D}_{\text{test}}$, \colorvar{\textit{model\_parameters}}: $\theta_{\text{global}}$, the last "outputs" of the Inner Optimization jobs: $\theta_{\text{Inner}}$ and the increments $\Delta \theta_{\text{Inner}}$, policy thresholds: $\gamma_t , \beta_t $, \cblue{TEST\_TYPE}
\Ensure A history of model parameters is available. (\cblue{TEST\_TYPE} = "B" requires the previous $\tau_c$ model parameters for the test condition) 
\If {\cblue{TEST\_TYPE} = "A"} \Comment{\color{brown}Check accuracy against other service providers~\cite{pmlr-v97-bhagoji19a}\color{black}}
    \State $\tilde{\theta} \leftarrow \theta_\text{global} + \Delta\theta^{\cblue{\text{SP\_NPUB}}}_{\text{Inner}} \ $
    \State $\tilde{\theta}_{G\setminus \text{\cblue{SP\_NPUB}}} \leftarrow \theta_\text{global} + \sum_{\text{npub}\in \text{\cblue{PROVIDERS}}}\Delta\theta^\text{npub}_{\text{Inner}} \ $
    \If {$\sum_{z\in \mathcal{D}_{\text{test}}}  \ell( \tilde{\theta}, z)   -\ell( \tilde{\theta}_{G\setminus \text{\cblue{SP\_NPUB}}} , z) > \gamma_t$}
    \State  \cblue{SP\_NPUB} is malicious
    \State \Return FAIL
    \EndIf
\ElsIf{\cblue{TEST\_TYPE} = "B"}    \Comment{\color{brown} Check progress over time (with time lag) \color{black}}
    \vspace{+0.1cm}
    \State $\theta^\tau \leftarrow \theta^\text{\cblue{SP\_NPUB}}_{\text{Inner at time $\tau$}}$, for all $\tau \in [t-\tau_c ,t]$ \Comment{\color{brown} For Outer: replace $ \theta^\text{\cblue{SP\_NPUB}}_{\text{Inner at time $\tau$}}$ with $ \theta^\text{\cblue{SP\_NPUB}}_{\text{global at time $\tau$}}$ \color{black}}
    \vspace{+0.1cm}
    \If {$\frac{1}{\tau_c +1}\sum^{t}_{\tau = t-\tau_c }\sum_{z\in \mathcal{D}_{\text{test}}}  \ell( \theta^{\tau}  , z )   > \beta_t$} \Comment{\color{brown} Check if a moving average of the loss decays \color{black}}
    \vspace{+0.1cm}
    \State \Return FAIL
    \EndIf
\EndIf
\State \Return PASS
\end{algorithmic}
\end{algorithm}

\subsection{Service Provider}\label{Section: SP}
Algorithm \ref{alg:service_provider} implements the routine for service providers. To advertise their services, service providers publish a discoverability event of kind 31990 (Event Type~\ref{Discover_Event}). Initially, the service providers wait until they receive a job request. Then they signal readiness through a job feedback event. Upon receiving any required payments, they execute the optimization routine, as requested by the customer. The optimization procedure uses one of two run options (FEDAVG or DiLoCo) and the task may be either Inner or Outer optimization. Depending on the run option the service providers calls the corresponding algorithmic routine (Algorithm \ref{alg:inner} or Algorithm~\ref{alg:outer}). Periodically, the service provider announces the status of the job through a Job Feedback event and provides  progress updates on the computation, for instance see Algorithm \ref{alg:inner}. After completing the optimization routine, service providers respond with a job result event, and expect any pending payment to be settled. If the customer does not complete payment(s) after a reasonable time period despite the delivery of accurate outputs, service providers may immediately stop accepting job requests from that customer. 
\begin{algorithm}
\caption{Service Provider}\label{alg:service_provider}
\begin{algorithmic}[1]
\Require \cblue{SP\_NPUB}: service provider public key, <computation information \& time>, <lightning address>
\Ensure Connection through websockets to all \colorvar{\textit{relays}} in \cblue{RELAYS} \Comment{\color{brown} Where events are published at\color{black}}
\State Publish an event \colorkind{(NIP-89) of kind $31990$} for discoverability \Comment{\color{brown} Event Type \ref{Discover_Event}\color{black}}
\While{\cblue{TRUE}}
\State \textbf{Routine:} Search for \colorkind{JOB REQUEST} for provider \cblue{SP\_NPUB} 
\If{\colorkind{JOB REQUEST} with appropriate KIND was found} \Comment{\color{brown} Event Type \ref{JRLabel_AI}\color{black}}
    \State Publish \colorkind{Job Feedback (KIND: 7000)} with \cblue{STATUS} = payment\_request \Comment{\color{brown} Event Type \ref{Job_Feedback_Event}\color{black}} 
        \State Wait until the initial payment is submitted, fetch and validate \colorkind{Payment Receipt (KIND: 9735)}         
        \If{payment = successful}
            \State Publish \colorkind{Job Feedback (KIND: 7000)} with \cblue{STATUS} = processing \Comment{\color{brown} Event Type \ref{Job_Feedback_Event}\color{black}}
        \Else
            \State Publish \colorkind{Job Feedback (KIND: 7000)} with \cblue{STATUS} = error
        \EndIf
        \State Fetch \colorvar{\textit{model\_parameters}}, \cblue{DATA\_SOURCES(SP\_NPUB)}, \cblue{ RUN\_OPTION}, \cblue{TASK}  from \colorkind{JOB REQUEST} 
        \If {\cblue{RUN\_OPTION} = \colorkind{FEDAVG} and \cblue{TASK} = \colorkind{INNER}}
            \State Call Inner\_Optimization( \colorvar{\textit{model\_parameters}}, \cblue{DATA\_SOURCES} , \colorkind{FEDAVG} )
        \ElsIf{\cblue{RUN\_OPTION} = \colorkind{FEDAVG} and \cblue{TASK} = \colorkind{OUTER}}
        \State Call Outer\_Optimization( \colorvar{\textit{model\_parameters}}, \cblue{DATA\_SOURCES} , \colorkind{FEDAVG} )
        \ElsIf{{\cblue{RUN\_OPTION} = \colorkind{DiLoCo}} and \cblue{TASK} = \colorkind{INNER}}
            \State Call Inner\_Optimization( \colorvar{\textit{model\_parameters}}, \cblue{DATA\_SOURCES} , \colorkind{DiLoCo} )
        \ElsIf{{\cblue{RUN\_OPTION} = \colorkind{DiLoCo}} and \cblue{TASK} = \colorkind{OUTER}}
         \State Call Outer\_Optimization( \colorvar{\textit{model\_parameters}}, \cblue{DATA\_SOURCES} , \colorkind{DiLoCo} )
        \EndIf
        \State Publish \colorkind{Job Feedback (KIND: 7000)} with \cblue{STATUS} = success \Comment{\color{brown} Event Type \ref{Job_Feedback_Event}\color{black}}
        \State Publish a \colorkind{JOB RESULT} event with "output" the result of the optimization \Comment{\color{brown} Event Type \ref{JRE}\color{black}}
    \EndIf
    \EndWhile

\end{algorithmic}
\end{algorithm}

\paragraph{Federated Averaging vs Distributed Low-Communication
Training of LLMs (DiLoCo)}
 The federated averaging algorithm~\cite{mcmahan2017communication} (FEDAVG) is a distributed and communication-efficient learning method for AI models trained on decentralized data. This decentralized computation is known as federated learning; a set of different devices learn a shared model by aggregating locally-computed updates. In our design, we consider the FEDAVG algorithm as one optimization approach. The customer provides a common model specification and the data to the service providers. The service providers have the role of different devices; they obtain a subset of the dataset and they train the shared model. Upon completion of the computation, they return the optimized model parameters and receive payment.

Although the FEDAVG is a successful approach for training neural networks and AI in a decentralized fashion, we also consider a variant that has numerous advantages over the classical approach. Specifically, the DiLoCo algorithm~\cite{douillard2023diloco} enables the training of LLMs under constrained communication. In fact, the inner part of optimization (jobs by service providers) requires the greatest amount of computation, while the frequency of new job requests is significantly reduced. This provides additional flexibility to our design by minimizing the number of job requests from the customer. Nevertheless, service providers can frequently provide information (e.g. job status, partial output) to the customer through the Job Feedback events. We proceed by presenting the implementation of the Inner Optimization and Outer Optimization routines for both FEDAVG and DiLoCo algorithms. 

\paragraph{Inner Optimization (Algorithm \ref{alg:inner})}
The Inner Optimization routine provides the implementation of the inner part of the computation. A service provider runs the inner part of FEDAVG (example SGD) or DiLoCo (AdamW~\cite[Algorithm 2]{loshchilov2017decoupled}). The run option is provided by the customer through the job request event. The algorithm returns the updated model parameter.
\begin{algorithm}
\caption{Inner\_Optimization}\label{alg:inner}
\begin{algorithmic}[1]
\Require 
\cblue{RUN\_OPTION}: <FEDAVG or DiLoCo>, \cblue{DATA:} \cblue{DATA\_SOURCES}, \colorvar{\textit{model\_parameters}}: $\theta_{\text{global}}$
\Ensure Connection through websockets to all \colorvar{\textit{relays}} in \cblue{RELAYS} \Comment{\color{brown} Where events are published at\color{black}}
\State $\theta \leftarrow \theta_{\text{global}}$
\If {\cblue{RUN\_OPTION} = FEDAVG} \Comment{\color{brown}Run the inner part of FEDAVG\color{black}}
    \State data $\leftarrow$ \cblue{DATA\_SOURCES(SP\_NPUB)}
    \State $\mathcal{B}$ $\leftarrow$ Split data into batches 
    
    \For {epoch $e \in (1, E)$} 
    \State Publish \colorkind{Job Feedback (KIND: 7000)} with \cblue{STATUS} = processing \Comment{\color{brown} Event Type \ref{Job_Feedback_Event}\color{black}}
        \For {batch $\boldsymbol{b}\in\mathcal{B}$}
        \State $\theta \leftarrow \theta - \eta \nabla \ell (\theta ; \boldsymbol{b}) $
        \EndFor
    \EndFor
\ElsIf{\cblue{RUN\_OPTION} = DiLoCo}    
\For {epoch $e \in (1, E)$}    
    \State Sample a batch $\boldsymbol{b}$ of data from \cblue{DATA\_SOURCES(SP\_NPUB)}
    \State Run AdamW($\theta , \nabla \ell (\theta , \boldsymbol{b})$) \Comment{\color{brown} AdamW~\cite[Algorithm 2]{loshchilov2017decoupled} \color{black}}
    \State Periodically publish \colorkind{Job Feedback (KIND: 7000)} with \cblue{STATUS} = processing \Comment{\color{brown} Event Type \ref{Job_Feedback_Event}\color{black}} 
    \State $\theta \leftarrow  $ AdamW($\theta , \nabla \ell (\theta , \boldsymbol{b})$) 
\EndFor
\EndIf
\State \Return $\theta$ \Comment{\color{brown} Returns $\theta^{(\cblue{\text{SP\_NPUB}})}_{\text{Inner}}$ \color{black}}
\end{algorithmic}
\end{algorithm}

\paragraph{Outer Optimization (Algorithm \ref{alg:outer})}
For the Outer Optimization routine, the customer can either run the outer optimization locally or assign it to a service provider. Then the customer (or a service provider) runs the outer part of FEDAVG (e.g., aggregation step) or DiLoCo~\cite[Nesterov Momentum]{douillard2023diloco,huo2020faster}. The run option is determined by the customer. The algorithm returns the updated model parameters.

\begin{algorithm}
\caption{Outer\_Optimization}\label{alg:outer}
\begin{algorithmic}[1]
\Require \cblue{RUN\_OPTION}: <FEDAVG or DiLoCo>, \cblue{DATA:} \cblue{DATA\_SOURCES}, \colorvar{\textit{model\_parameters}}: $\theta_{\text{global}}$, the last "outputs" of the Inner Optimization jobs: $\theta_{\text{Inner}}$ 
\Ensure Connection through websockets to all \colorvar{\textit{relays}} in \cblue{RELAYS} \Comment{\color{brown} Where events are published at\color{black}}
\If {\cblue{RUN\_OPTION} = FEDAVG} \Comment{\color{brown}Run the aggregation (outer) part of FEDAVG\color{black}}
    \State  $\theta_{\text{global}} = \frac{1}{\cblue{\text{NUM\_PR}}} \sum^{\cblue{\text{NUM\_PR}}}_{k=1} \eta_k \times \theta^{(k)}_{\text{Inner}}$ \Comment{\color{brown} For some weights $\eta_k$ \color{black}}
\ElsIf{\cblue{RUN\_OPTION} = DiLoCo}    \Comment{\color{brown} Run the Outer part of DiLoCo~\cite{douillard2023diloco} \color{black}}
\State  $\Delta \theta_{\text{Outer}}  = \frac{1}{\cblue{\text{NUM\_PR}}} \sum^{\cblue{\text{NUM\_PR}}}_{k=1} \eta_k \times ( \theta_{\text{global}} - \theta^{(k)}_{\text{Inner}})$ 
\State Run Nesterov Momentum:  \Comment{\color{brown} See FedMom~\cite[Algorithm 3]{huo2020faster} \color{black}}
\State (Periodically) publish \colorkind{Job Feedback (KIND: 7000)} with \cblue{STATUS} = processing \Comment{\color{brown} Event Type \ref{Job_Feedback_Event}\color{black}} 
\State $\theta_{\text{global}}\leftarrow $ Nesterov\_Momentum ($\theta_{\text{global}}$ , $\Delta \theta_{\text{Outer}}$)
\EndIf
\State \Return $\theta_{\text{global}}$
\end{algorithmic}
\end{algorithm}





\section{Proof-of-Concept Implementation}

To validate the FEDSTR protocol design, we developed a working proof-of-concept implementation demonstrating federated learning over the NOSTR network with cryptographic integrity guarantees. The primary contribution of this implementation is the demonstration that \textbf{(1)} the NOSTR protocol can successfully orchestrate distributed federated learning, \textbf{(2)} model validation ensures model integrity across untrusted relays, and \textbf{(3)} the system operates on public, decentralized infrastructure without trusted intermediaries.

\subsection{Implementation Overview}

The proof-of-concept implements the complete FEDSTR protocol flow as described in Section 7, including:

\begin{itemize}
    \item Customer coordinator managing federated learning rounds 
    \item Data Vending Machines (DVMs) as federated learning service providers 
    \item Communication over public NOSTR relays: All events are cryptographically signed using Schnorr signatures and propagated through multiple public NOSTR relays, demonstrating the protocol's decentralized nature.
    \item SHA-256 cryptographic hash verification for model integrity at every exchange
    \item Validation of DVM outputs 
    \item Model storage with hash-based content addressing
\end{itemize}

\textbf{Note:} Similarly to exchanging notes (messages on NOSTR), exchanging model parameters also satisfies the following:

\begin{itemize}
    \item Relays cannot modify model parameters without detection
    \item Man-in-the-middle attacks are cryptographically prevented
    \item The protocol achieves model integrity without trusted intermediaries
    \item Hash verification adds negligible computational overhead 
\end{itemize}

\textbf{Code Availability:} The proof-of-concept is open source and available on GitHub~\cite{fedstr_github}.

\subsection{Storage Architecture}

The implementation supports multiple storage backends for model parameters.
\vspace{+0.2cm}

\noindent \textbf{Current Implementation:} Local filesystem storage (\texttt{file://} URLs) with SHA-256 hash-based filenames. Models are serialized to raw bytes ($\sim$400KB) and stored at \texttt{/tmp/fedstr\_models/model\_<hash>.bin}.
\vspace{+0.2cm}

\noindent \textbf{Example of Storage Operations:}
\begin{itemize}
    \item \textbf{Upload:} DVM serializes model $\rightarrow$ computes hash $\rightarrow$ saves to storage $\rightarrow$ publishes URL + hash
    \item \textbf{Download:} Customer retrieves URL from JobResult $\rightarrow$ downloads bytes $\rightarrow$ verifies hash $\rightarrow$ deserializes model
\end{itemize}

\noindent\textbf{Extensibility:} The storage layer is abstracted to support future backends:
\begin{itemize}
    \item HTTP servers for remote storage (Tested already with private storage service)
    \item IPFS for content-addressed distributed storage  
    \item Blossom servers (NOSTR-native file storage)
    \item NIP-94 file metadata events (see Section \ref{sec:nip94})
\end{itemize}

\noindent All the coordination and ML operations are \textit{storage-agnostic} -- integrity is guaranteed regardless of storage backend, as long as the model parameters' hash matches.

\subsection{Current Limitations}

The proof-of-concept demonstrates core protocol functionality but intentionally simplifies certain components for clarity:

\begin{itemize}
    \item \textbf{Payments:} Dummy Lightning Network payment implementation (no actual bitcoin transfers). Events include bolt11 invoice fields, but payment verification is stubbed. Full NIP-57 integration is planned.
    
    \item \textbf{DVM Identity:} DVMs generate fresh keypairs on startup. Production systems should use persistent keypairs stored securely.
    
    \item \textbf{Deployment:} Single-machine simulation (DVMs run as separate processes). Protocol is designed for multi-machine deployment, which remains to be tested at scale.
    
    \item \textbf{Reputation:} No reputation system implemented. Trust relies solely on cryptographic verification and validation.
    
    \item \textbf{Storage:} Local file storage and HTTP servers for remote storage. Decentralized cloud storage integration (Blossom, IPFS) is implemented but not yet tested in production.
\end{itemize}

These limitations do not affect the \textit{core protocol validation} -- they represent engineering work required for production deployment.

\subsection{Future Work -- NIP-94 Integration}\label{sec:nip94}

While the current proof-of-concept embeds model URLs directly in JobResult events (Kind 6000), future versions could leverage NIP-94 (File Metadata) \cite{nip94} for improved model discoverability and relay indexing.

\subsubsection{NIP-94 Protocol Overview}

NIP-94 defines a standardized event type (Kind 1063) for publishing file metadata to NOSTR relays. Files are described with tags including URL, MIME type, SHA-256 hash, size, and optional metadata (thumbnails, descriptions, etc.). Relays can index these events, enabling file discovery and categorization.

\subsubsection{Potential Benefits for FEDSTR}

\begin{enumerate}
    \item \textbf{Standardization:} Trained models published as standard NOSTR file events, interoperable with other NOSTR file applications.
    
    \item \textbf{Discoverability:} Relays can index trained models by dataset, accuracy, model architecture, etc., enabling "model marketplace" discovery.
    
    \item \textbf{Rich Metadata:} Attach training metrics, model architecture details, dataset information, and visualization thumbnails to model files.
    
    \item \textbf{Reusability:} Same trained model can be referenced from multiple JobResults without event duplication.
    
    \item \textbf{Relay Indexing:} Specialized "model relays" could index only Kind 1063 events, creating curated model repositories.
\end{enumerate}

\subsubsection{Proposed Implementation}

Under a NIP-94 integration, the protocol flow would be modified as follows:

\begin{enumerate}
    \item DVM completes training and uploads model to storage (unchanged)
    \item DVM publishes NIP-94 file metadata event (Kind 1063):
\end{enumerate}

\begin{center}
\begin{minipage}{0.8\textwidth}
\begin{verbatim}
{
  "kind": 1063,
  "tags": [
    ["url", "https://blossom.server/model_abc123.bin"],
    ["x", "abc123..."],  // SHA-256 hash
    ["m", "application/octet-stream"],
    ["size", "409245"],
    ["alt", "FEDSTR MNIST Round 3 - 97.28% accuracy"]
  ],
  "content": "Federated learning model trained on MNIST"
}
\end{verbatim}
\end{minipage}
\end{center}

\begin{enumerate}
    \setcounter{enumi}{2}
    \item DVM publishes JobResult (Kind 6000) referencing the NIP-94 event:
\end{enumerate}

\begin{center}
\begin{minipage}{0.8\textwidth}
\begin{verbatim}
{
  "kind": 6000,
  "tags": [
    ["job_id", "abc123"],
    ["e", "<nip94_event_id>", "wss://relay.damus.io"],
    ["loss", "0.0875"],
    ["accuracy", "0.9728"]
  ]
}
\end{verbatim}
\end{minipage}
\end{center}

\begin{enumerate}
    \setcounter{enumi}{3}
    \item Customer retrieves JobResult, follows the event reference to NIP-94 event, downloads model from URL, verifies hash (unchanged verification protocol)
\end{enumerate}





\subsubsection{Recommendation}

We recommend the following strategy:

\begin{itemize}
    \item \textbf{Private FL Applications:} Use direct URL embedding in JobResult (current approach). Simpler, faster, sufficient for closed federations.
    
    \item \textbf{Public Model Sharing:} Use NIP-94 dual-event approach. Enables discoverability and marketplace features.
    
    \item \textbf{Hybrid:} Publish both - embed URL in JobResult for efficiency, also publish NIP-94 for discoverability.
\end{itemize}

The current protocol design is \textit{compatible with all three approaches} -- the choice can be made at deployment time based on application requirements. Future versions of the PoC will include NIP-94 support as a configurable option.

\subsection{Deployment Considerations}

The proof-of-concept demonstrates that FEDSTR can operate on existing NOSTR infrastructure without modification:

\begin{itemize}
    \item \textbf{Public Relays:} Successfully used relay.damus.io and nos.lol (free, public, no registration required)
    
    \item \textbf{No Special Infrastructure:} No dedicated "FEDSTR relays" needed -- standard NOSTR relays handle all event types
    
    \item \textbf{Client Compatibility:} DVMs and customers are standard NOSTR clients -- can use existing libraries (nostr-sdk, NDK, etc.)
    
    \item \textbf{Incremental Deployment:} Can scale to multiple DVMs without protocol changes
    
    \item \textbf{Self-Hosted Infrastructure (Optional):} Users have the option to run their own relays and storage servers for maximum privacy and control. By self-hosting, organizations can:
    \begin{itemize}
        \item Monitor all coordination and events in real-time
        \item Guarantee data locality and regulatory compliance
        \item Optimize network performance (reduce latency, increase bandwidth)
        \item Maintain complete privacy (events never leave private infrastructure)
        \item Control relay policies (access control, rate limiting, etc.)
    \end{itemize}
    Critically, \textit{self-hosted deployments use the exact same protocol} as public deployments -- no modifications required. A private FEDSTR federation can seamlessly transition to public relays or vice versa by simply changing relay URLs, demonstrating the protocol's flexibility.
\end{itemize}

This demonstrates a key advantage of building on NOSTR: the protocol inherits the existing decentralized infrastructure without requiring specialized setup. Simultaneously, it supports fully private deployments for organizations with strict privacy or performance requirements.

\subsection{Summary of Proof-of-Concept Contributions}

The PoC validates the following key claims:

\begin{enumerate}
    \item \textbf{Protocol Correctness:} The FEDSTR event flow (JobRequest $\rightarrow$ JobFeedback $\rightarrow$ JobResult) successfully orchestrates federated learning across multiple DVMs
    
    \item \textbf{Cryptographic Integrity:} Cryptographic verification ensures integrity across untrusted relay infrastructure
    
    \item \textbf{Decentralization:} The system operates on public NOSTR relays with no trusted intermediaries
    
    
    \item \textbf{Infrastructure Reuse:} FEDSTR works on existing NOSTR infrastructure without modifications
\end{enumerate}

These results demonstrate that \textbf{decentralized federated learning is viable on the NOSTR protocol}, validating the core thesis of this work. Complete implementation details and source code are available at 
\url{https://github.com/ConstantinosNikolakakis/Fedstr}.

\section{Conclusion}
The NOSTR protocol represents a paradigm shift for reliable and decentralized social web applications. In this work, we introduced a design and protocol flow for LLM and AI training on decentralized marketplace system, which builds upon NOSTR. The fast communication through websockets and the flexibility of protocol development enables the possibility of a competitive, open, and censorship-resistant computation network. In this network, users (customers) provide data and payment, and receive a trained AI model. The integrated payment protocol and the Lightning network offer fast, reliable payments with virtually zero transaction fees and without a trusted party. As a result, service providers can maximize their profit and compete within an open market. Further, this payment system supports microtransactions with instant settlement, which serves as an effective anti-spam mechanism, provides a reliable form of handshaking between customers and service providers, enables on-the-fly payments upon receiving partial computational results, and supports payments upon validation of the final output.

We validated the FEDSTR protocol design through a working proof-of-concept 
implementation demonstrating successful federated learning over public and private NOSTR 
relays with cryptographic integrity guarantees. The implementation proves that:
(1) The NOSTR protocol can successfully orchestrate distributed federated 
learning without trusted intermediaries,
(2) Cryptographic hash verification ensures model integrity across untrusted 
relay infrastructure,
(3) The protocol operates on existing public NOSTR infrastructure without 
requiring specialized setup.
These results demonstrate the viability of FEDSTR for real-world decentralized 
federated learning applications, validating the core thesis that the NOSTR 
protocol provides a suitable foundation for AI training, inference marketplaces, and agentic AI systems.

We considered a simple but effective mechanism for validation of the output, by periodically checking the progress of the trained model through a validation dataset. However, alternative solutions for verifying computational accuracy could also be considered. Such robust approaches can be designed using smart contracts and the BitVM mechanism~\cite{bitvm}, which enables verification of any computable function. Future work may also explore the inclusion of verification mechanisms and reputation systems, as well as application-specific protocol flow variations supporting alternative AI algorithms—such as DeepSpeed~\cite{rasley2020deepspeed}, Zero Redundancy Optimizer~\cite{rajbhandari2020zero}, and distributed LLM deployment and inference~\cite{fan2023fate,lu2023exploring,tang2023fusionai,agrawal2023sarathi,shen2023efficient,spector2023accelerating,cai2024medusa,del2023skipdecode}.

\bibliographystyle{unsrt}
\bibliography{references}

\end{document}